\documentclass{article}
\usepackage[utf8]{inputenc}
\usepackage{graphicx}
\usepackage{amsmath}
\usepackage{amssymb}
\usepackage{geometry}
\usepackage{hyperref}

\title{An Electromagnetic Framework for the Deployment of Reconfigurable Intelligent Surfaces to Control Massive MIMO Channel Characteristics}
\author{Debdeep Sarkar, Said Mikki and Yahia Antar}
\date{November 2019}

\begin{document}

\maketitle
\footnotetext[1]{Debdeep Sarkar \textit{(Corresponding Author)} and Yahia Antar are with the Royal Military College, PO Box 17000, Station Forces Kingston, Ontario, K7K 7B4, Canada (Emails: \textit{debdeep1989@gmail.com}; antar-y@rmc.ca).}
\footnotetext[2]{Said Mikki is with University of New Haven, West Haven, Connecticut, 300 Boston Post Rd, 06516, USA (Email: said.m.mikki@gmail.com).}
\footnotetext[3]{$\copyright$20XX IEEE.  Personal use of this material is permitted.  Permission from IEEE must be obtained for all other uses, in any current or future media, including reprinting/republishing this material for advertising or promotional purposes, creating new collective works, for resale or redistribution to servers or lists, or reuse of any copyrighted component of this work in other works.”}

\begin{abstract}
In this paper, we deploy a full-wave FDTD paradigm to investigate the effect of reconfigurable intelligent surface (RIS) -- switchable frequency-selective surfaces (FSS) -- on generic massive MIMO uplink channel's eigenspace structure. We place an RIS based on two switchable FSS layers in the vicinity of a 64-element massive MIMO base-station (BS) array, serving a cluster of four fixed user equipment (UE) units. Utilizing an electromagnetic tool based on time-averaged Poynting flow developed recently by the authors, we demonstrate how the illumination of BS-array aperture can be controlled by the intentional deployment of various switching states in the RIS placed near the BS. We show that such supplementary RIS structures may assist the wireless link engineer in deterministically ``customizing'' the uplink channel behaviour by selectively enhancing/suppressing certain channel eigenvalues.  
\end{abstract}

\section{Introduction}
Massive MIMO (multiple-input multiple-output) systems have demonstrated promising performance in terms of spectral, energy and hardware efficiency, which has established it as one of the key underlying technologies for the sub-6 GHz and mm-wave wireless networks (5G/6G) of future generation \cite{mimo_book2}-\cite{mmimo_review}. Currently active research on massive MIMO technology is seeing the emergence of futuristic concepts like ``extremely large aperture arrays'' (ELAAs), ``cell-free massive MIMO'' and ``large intelligent surface'' (LIS), pushing the engineers to look into the spatial correlation and channel-modelling aspects from an advanced electromagnetic perspective  \cite{emil_future}-\cite{emil_future2}. Traditionally, researches on massive MIMO channel modelling have been either experimental/empirical \cite{Weng}-\cite{she}, or based on simplified assumptions like simplified uncorrelated Rayleigh fading models or ray-tracing tracing approach. With a large number of antennas being accommodated both in the base-stations (BS) as well as user-equipments (UEs), it has become crucial for correlation analysis/channel-modelling to extract the complete \textit{electromagnetic} (EM) knowledge by incorporating effects of the impinging spherical wave-fronts of near-field illumination, port-to-port mutual coupling as well as relative BS-element polarization \cite{Mikki_book}. The problem of far-field spatial correlation analysis in MIMO systems having multiple interacting elements with arbitrary relative pattern and polarization, can be analytically handled by utilizing infinitesimal dipole models (IDMs) in conjunction with ``cross-correlation Green's functions'' (CGFs) \cite{mikki_cgf_2015}-\cite{idm_cgf_mmimo_2019}. However, it is important to utilize full-wave EM simulation techniques like the finite-difference time-domain (FDTD) method in channel-modelling, in order to properly account for the port-to-port mutual coupling, spherical wave-front illumination and near-field correlation scenarios \cite{yee}-\cite{nf_corr_awpl}. 

\begin{figure}[htbp]
\begin{center}
\includegraphics[width=12cm]{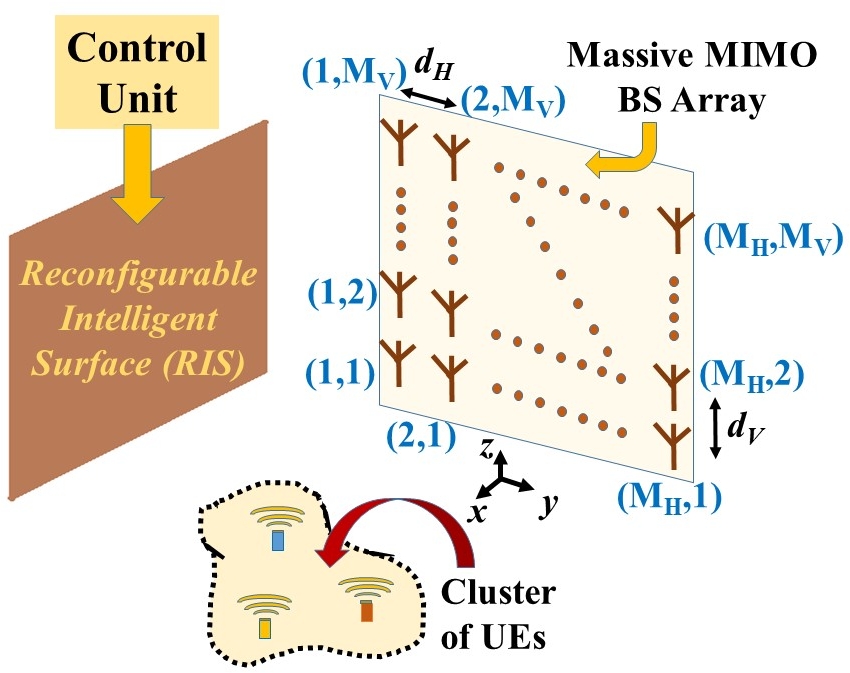}
\caption{Vision for ``wireless 2.0'' \cite{Hu_2}-\cite{ntontin}: Schematic diagram showing use of reconfigurable intelligent surfaces (RISs) in conjunction with massive MIMO base-station (BS) arrays serving a dense cluster of user-equipments (UEs) in a ``smart-radio'' environment. The massive MIMO configuration is a uniform rectangular array (URA) of $M=M_{V}M_{H}$ elements, with respective inter-element spacings $d_{H}$ and $d_{V}$ along $y$ and $z$ directions \cite{idm_cgf_mmimo_2019}.} \label{fig0}
\end{center}
\vspace{-10pt}
\end{figure}

At this juncture, it is interesting to note that the so-called ``massive MIMO 2.0'' for smart wireless environments envisions the application of switchable/tunable frequency selective surfaces (FSSs), often termed as ``reconfigurable intelligent surface (RIS)'', as an alternative to active relays \cite{emil_future}-\cite{emil_future2}, \cite{Hu_2}-\cite{ntontin}. Such RIS is typically based on periodically arranged low-cost scatterers/reflectors\footnote{This is conceptually similar to classic reflectarrays \cite{Hum}, or the idea of coding meta-surfaces \cite{akbari}.} and is capable of electromagnetically controlling the channel properties, by remaining in the vicinity of the massive MIMO base-stations \cite{Hu_2}-\cite{ntontin}. Fig. \ref{fig0} shows the topology of the massive MIMO system where an RIS is kept in the vicinity of the BS massive MIMO array for efficiently communicating with a dense cluster of UEs. As recently emphasized upon in \cite{mmimo_FDTD_FSS_awpl}, full-wave FDTD is extremely relevant in channel modelling for massive MIMO systems equipped by RIS, since unlike conventional ray-tracing (RT) technique, FDTD inherently accounts for the near-field coupling and EM interaction effects of: (i) the individual UE/BS antenna elements, (ii) the closely spaced RIS unit-cells, and (iii) the RIS with the massive MIMO BS-array/UEs. 

However, in \cite{mmimo_FDTD_FSS_awpl} we only focus on the channel-manipulation by varying the periodicity of \textit{passive} FSS, but do not demonstrate the channel eigenspace manipulation via \textit{reconfigurability/switchability} in the FSS. Therefore in this paper, we first determine the channel matrix for the UE-to-BS massive MIMO uplink ($\mathbf{H}=\mathbf{H}_{\text{uplink}}$) following the FDTD-based EM-approach of \cite{mmimo_FDTD_FSS_awpl}, and consequently study the impact of RIS on the eigan-value distribution of the Gram matrix $\mathbf{G}=\mathbf{H}^{H}\mathbf{H}$.  

\section{Massive MIMO Channel eigenspace Manipulation using RIS}
\subsection{FDTD Computational Setup: Performance without RIS}
We consider FDTD framework based on Cartesian coordinate system with cubical spatial grids ($\Delta x=\Delta y=\Delta z=\Delta = 12$ mm) and time-step $\Delta t=18.4$ ps which satisfies the CFL criterion \cite{ds_tap_1}, \cite{mmimo_FDTD_FSS_awpl}. The total FDTD computational volume is $145\Delta \times 120\Delta \times 170\Delta$, with $10\Delta$-long Berenger's perfectly matched layers (PML) padded along $x,y$ and $z$ directions. 

\begin{figure}[htbp]
\begin{center}
\includegraphics[width=12cm]{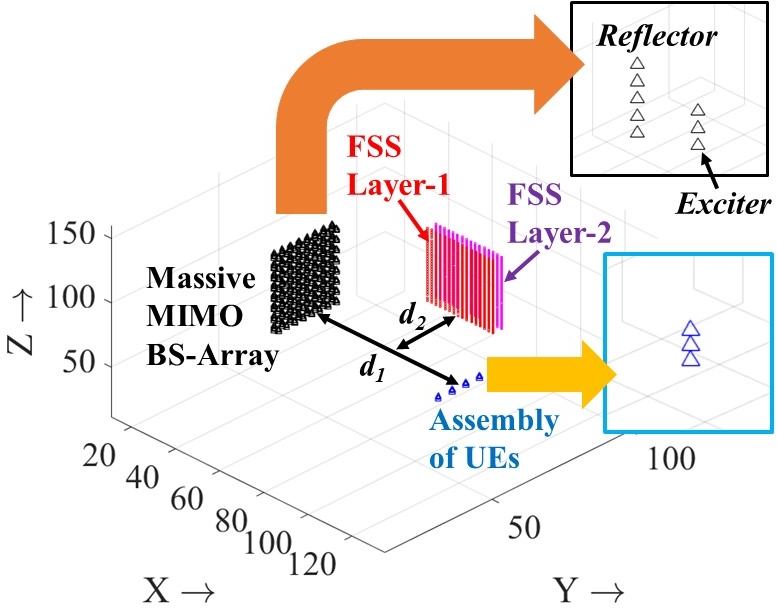}
\caption{Full-wave FDTD simulation setup comprising of: (i) 64-element massive MIMO BS-array consisting of directional elements (reflector-backed short dipoles), (ii) an assembly of 4 UEs having omni-directional elements (short dipoles) and (iii) an RIS with two FSS layers. The zoomed views represent the infinitesimal dipole model (IDM) representation of UE and BS antenna elements. The parameters $d_{1}=70\Delta$ and $d_{2}=18\Delta$, where $\Delta$ is the cubical FDTD grid-size.} \label{fig1}
\end{center}
\vspace{-10pt}
\end{figure}

Fig. \ref{fig1} demonstrates the FDTD simulation paradigm for the complete system under consideration, having: (i) massive MIMO BS-antenna array of $M=64$ elements ($M_{V}=M_{H}=8$, $d_{V}=8\Delta$ and $d_{H}=3\Delta$, see Fig. \ref{fig0}) along $yz$-plane, (ii) assembly of $N=4$ UE antennas along $y$-axis with uniform spacing of $5\Delta$, and (iii) RIS consisting of two FSS layers along the $xz$-plane. The BS antenna elements are reflector-backed short-dipoles having end-fire directional patterns, while the UE antennas are short-dipoles with omni-directional radiation, both having the same polarization (along $z$-direction, see Fig. \ref{fig1}). Both the UE antenna and the BS-exciter elements are modelled by three infinitesimal dipoles (IDs) \cite{mmimo_FDTD_FSS_awpl}, while the reflector for the BS-element consists of five IDs (Fig. \ref{fig1}). The UE-assembly along $xy$-plane is kept at $z=z_{0}-20\Delta$ (slightly offset downward), where $z_{0}=85\Delta$ lies in the mid-way of the computational volume (see Fig. \ref{fig1}). 

\begin{figure}[htbp]
\begin{center}
\includegraphics[width=12cm]{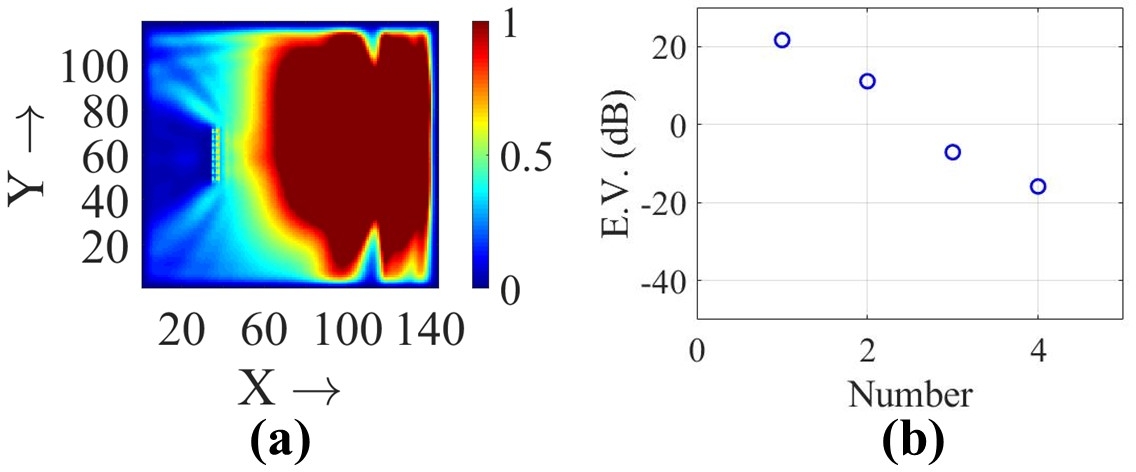}
\caption{(a) Normalized $\left|\mathbf{S}_{\text{avg}}\right|$ distribution (see \eqref{savg}) along  $xy$-plane ($z=z_{\text{center}}-20\Delta$, where the UEs are located), for the system of Fig. \ref{fig1} without RIS. Here UE-4 is chosen as the excited element. (b) Eigenvalue distributions for $\mathbf{G}=\mathbf{H}^{H}\mathbf{H}$, where $\mathbf{H}$ is the complete FDTD-computed channel matrix for the reference system without RIS.} \label{fig2}
\end{center}
\vspace{-10pt}
\end{figure}

Note that we deal with the UE-to-BS uplink channel $\mathbf{H}=\mathbf{H}_{\text{uplink}}$ for the time-being, implying that the BS-array operates in the receiving mode, i.e. all the exciter dipoles are maintained at matched load termination \cite{mmimo_FDTD_FSS_awpl}. The UEs are excited by a modulated Gaussian pulse $v_{\text{in}}(t)$ via their Delta feed-gap: 
\begin{equation}\label{vin_modgauss}
    v_{\text{in}}(t)= \exp\left[-\frac{(t-t_{0})^{2}}{2\sigma_{0}^{2}} \right]\sin\left(2\pi f_{0}t \right) ,
\end{equation}
where modulating signal frequency $f_{0}$ is 2.5 GHz, with parameters $t_{0}=70\Delta t$ and $\sigma_{0}=15\Delta t$. When one UE is excited, the others are terminated by matched load. To make the transient effects die down completely, the FDTD time-marching simulation is run for $T_{\text{max}}=400\Delta t$. Note that, the distance $d_{1}$ along $x$-axis between the UE-assembly and the massive MIMO BS-array is chosen at $70\Delta = 7\lambda_{0}$, where $\lambda_{0}=c/f_{0}$ and $c$ is the speed of EM wave in free-space.  

To visualize the EM-wave propagation from the UEs to the illuminated massive MIMO BS-arrays, we plot the 2D profiles of time-averaged Poynting vector magnitude $\left|\mathbf{S}_{\text{avg}}\right|$ given by:
\begin{equation}\label{savg}
    \mathbf{S}_{\text{avg}}=\mathbf{S}_{\text{avg}}\left(\mathbf{r}\right)=\frac{1}{T_{\text{max}}} \int_{0}^{T_{\text{max}}}\mathbf{S}\left(\mathbf{r},t \right) dt, 
\end{equation}
where the Poynting vector $\mathbf{S}=\mathbf{S}\left(\mathbf{r},t \right)=\mathbf{E}\left(\mathbf{r},t \right)\times\mathbf{H}\left(\mathbf{r},t \right)$ is computed via standard grid-interpolation techniques using local electric and magnetic fields (i.e. $\mathbf{E}\left(\mathbf{r},t \right)$ and $\mathbf{H}\left(\mathbf{r},t \right)$) generated by FDTD simulations \cite{mmimo_FDTD_FSS_awpl}. Fig. \ref{fig2}(a) shows the normalized $|\mathbf{S}_{\text{avg}}|$-distribution along the $xy$-plane ($z=z_{\text{center}}-20\Delta$), when UE-4 is excited by $v_{\text{in}}(t)$, in the system of Fig. \ref{fig1} without considering any RIS. 

Now to compute the frequency-domain $M \times N$ channel matrix $\mathbf{H}$ at $f_{0}$, we first excite the UE antenna-$j$ (where $j=1,2,\ldots,N$), then record the channel response signals $h_{ij}(t)$ at the BS antenna-$i$ (where $i=1,2,\ldots,M$), and finally Fourier transform the temporal signals $h_{ij}(t)$. Fig. \ref{fig2}(b) shows the eigenspace of the resulting $N \times N$ Gram-matrix $\mathbf{G}=\mathbf{H}^{H}\mathbf{H}$ (superscript $H$ indicates Hermitian) for the UE-array to-BS uplink without any RIS. Fig. \ref{fig2}(b) confirms that there are $N=4$ dominant eigen-channels and the corresponding eigenvalues are gradually decreasing in order.

\begin{figure}[htbp]
\begin{center}
\includegraphics[width=12cm]{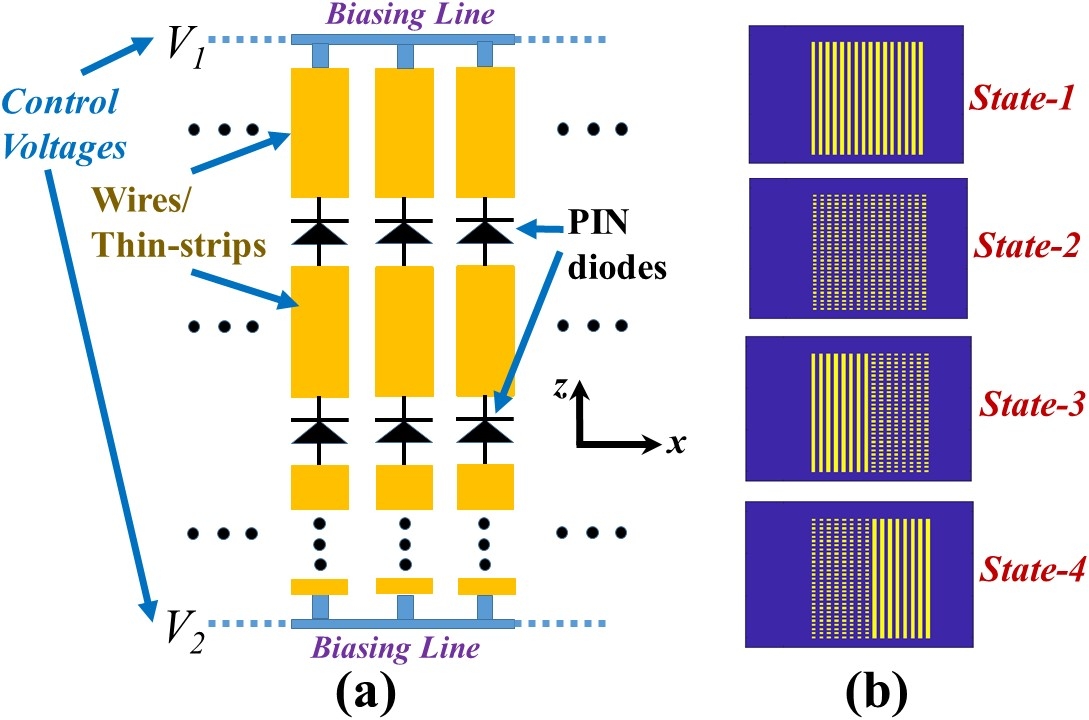}
\caption{(a) Possible practical implementation scheme for the switchable FSS-layers constituting the RIS structure of Fig. \ref{fig1}. (b) Four switching states of the FSS layers, which can be realized by use of PIN diodes and suitable biasing mechanism. In the FDTD code, we use connected wires for switch ON, and disconnected wires for switch OFF, for EM simulation purpose.} \label{fig3}
\end{center}
\vspace{-10pt}
\end{figure}

\subsection{Controlling Channel Eigenspace by Reconfigurability in FSS Layers of the RIS}
The RIS as shown in Fig. \ref{fig1} is typically mounted in a wall to provide additional paths to the UE-emitted signal for illuminating the BS massive MIMO array aperture. In our analysis we consider the RIS two be consisting of two FSS layers, each consisting of $z$-directed thin PEC-wires having periodicity $2\Delta$ along $x$-direction. The location of RIS relative to the UE-assembly and BS-array is specified by the parameters $d_{1}$ and $d_{2}$ (see Fig. \ref{fig1}). To realize the reconfigurability practically\footnote{One can find discussions on practical realization of such structures by utilizing thin microstrip-lines printed on low-loss substrates or by capillaries filled with liquid-metal technology (See \cite{bayatpur}-\cite{sghosh}).}, we need to load the wires with PIN diodes and control their ON-OFF condition by proper biasing circuitry (as depicted in the possible schematic of Fig. \ref{fig3}(a)). 

\begin{figure}[htbp]
\begin{center}
\includegraphics[width=12cm]{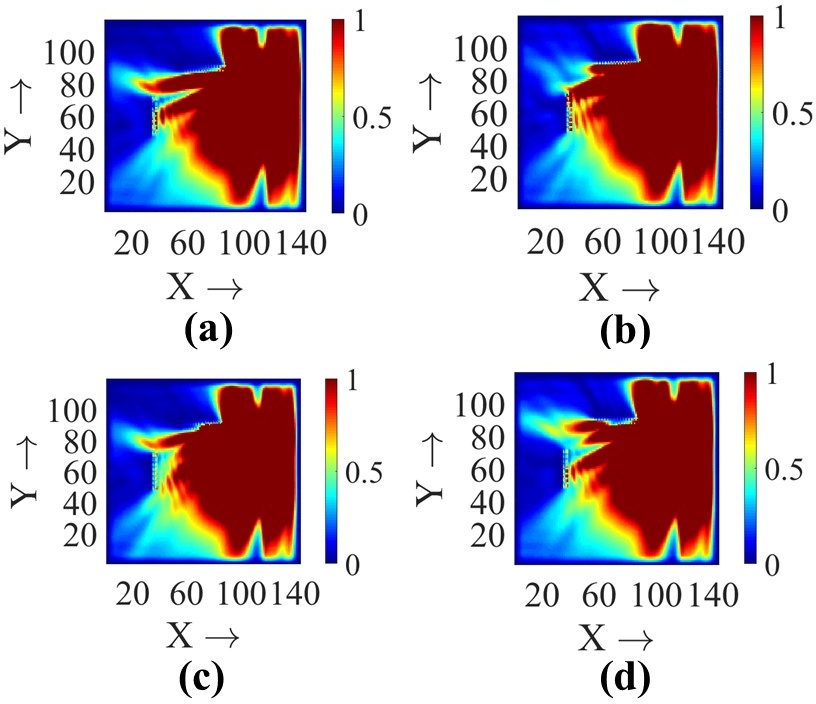}
\caption{Using UE-4 as the excited element, normalized $\left|\mathbf{S}_{\text{avg}}\right|$ distribution along $xy$-plane ($z=z_{\text{center}}-20\Delta$, see Fig. \ref{fig2}), for different switching states of FSS layer-1: (a) State-1, (b) State-2, (c) State-3 and (d) State-4. The FSS layer-2 (see Fig. \ref{fig1}) is always kept at State-1 as depicted in Fig. \ref{fig3}(b).} \label{fig4}
\end{center}
\vspace{-10pt}
\end{figure}

\begin{figure}[htbp]
\begin{center}
\includegraphics[width=10cm]{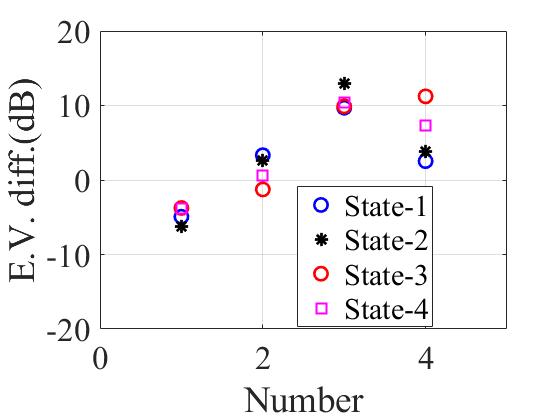}
\caption{Difference in Eigenvalues for  
$\mathbf{G}=\mathbf{H}^{H}\mathbf{H}$ (where $\mathbf{H}$ is the uplink channel matrix) for the RIS-based systems (using State-1 to State-4 for FSS layer-1, see Fig. \ref{fig5}), with respect to the reference system (i.e. without RIS).} \label{fig5}
\end{center}
\vspace{-10pt}
\end{figure}

In Fig. \ref{fig3}(b), we demonstrate four possible switching states where the FSS-layers can be re-configured by properly controlling the biasing voltages $V_{1}$ and $V_{2}$. Two important points must be mentioned here: (i) many other states apart from the ones (1 to 4, as shown in Fig. \ref{fig3}(b)) can be realized, but are not taken up for this particular work, (ii) in the FDTD simulations, we simply use the EM-aspects only and emulate the ON- and OFF-states of the switches as connected and disconnected PEC wires respectively. For our analysis, we choose FSS-2 in Fig. \ref{fig1} in a fixed state-1 (see Fig. \ref{fig3}), while vary the FSS-1 in four states 1 to 4, and re-compute the $M \times N$ channel matrix $\mathbf{H}$ for all scenarios using same method discussed in the previous section. Fig. \ref{fig4} illustrates how these various FSS-states leads to different $\mathbf{S}_{\text{avg}}$-configurations, and illuminates the BS massive MIMO-array aperture in different ways. Interestingly, when both FSS-1 and FSS-2 are in State-1 of Fig. \ref{fig3}(b), the RIS would behave like a simple PEC wall for the frequency of interest. However, it should be remembered that the presence of a simple PEC wall instead of an RIS provides a \textit{fixed} reflected beam, while variable FSS switching states in the RIS provide an opportunity to obtain different illumination-profiles of the BS-array for the same UE location.  

To quantify the effects of RIS we compute eigenvalue difference (similar to \cite{mmimo_FDTD_FSS_awpl}), by subtracting the  eigenvalues of $\mathbf{G}=\mathbf{H}^{H}\mathbf{H}$ without the RIS (i.e. reference case, see Fig. \ref{fig2}) from the eigenvalues of $\mathbf{G}=\mathbf{H}^{H}\mathbf{H}$ with the RIS-wall. It is clearly observed that the lower eigenvalues are enhanced in value for some specific switching configurations and State-3 in our analysis shows the most promising effect. 

\section{Conclusion}
Considering a fixed assembly of 4 UEs, the present paper uses a full-wave FDTD simulation approach in uplink channel modelling for an RIS-enabled 64 element massive MIMO BS-array of reflector-backed dipoles. The eigenvalue difference results of Fig. \ref{fig5} make it evident that the methodology based on switchable FSS or RIS has immense potential in tailoring/``customizing'' the channel eigenspace (i.e. suppressing/enhancing certain eigen-channels), even for a line-of-sight (LOS) type propagation scenario.

Note that, here we consider the UE and BS antenna-elements to beam same $z$-directed polarization, but other possible combinations can be easily explored in the FDTD simulation paradigm. Furthermore, we only consider only a few possible switching states (see Fig. \ref{fig3}) of top FSS-layer in this work. Future works will explore the application of efficient optimization routines for a particular application scenario, that can choose the desired RIS states from the multitude of possible realizable switching combinations.


\begin{thebibliography}{99}
\bibitem{mimo_book2} T. L. Marzetta, E. G. Larsson, H. Yang, and H. Q. Ngo, \textit{Fundamentals of Massive MIMO}, Cambridge, U.K.: Cambridge Univ. Press, 2016.






\bibitem{emil1} E. Bjornson, J. Hoydis and L. Sanguinetti, \textit{Massive MIMO Networks: Spectral, Energy, and Hardware Efficiency}, Foundations and Trends in Signal Processing: vol. 11, no. 3-4, pp. 154–655, 2017. DOI: 10.1561/2000000093.





\bibitem{mmimo_review} E. Bjornson, L. V. der Perre, S. Buzzi, and E. G. Larsson, ``Massive MIMO in sub-6 GHz and mm-Wave: Physical, practical, and use-case differences,'' \textit{IEEE Wireless Communications}, vol. 26, no. 2, pp. 100-108, 2019.

\bibitem{emil_future} E. Bjornson, L. Sanguinetti, H. Wymeersch, J. Hoydis and T. L. Marzetta, ``Massive MIMO is a Reality - What is Next? Five Promising Research Directions for Antenna Arrays,'' arXiv:1902.07678v1 [eess.SP], 11 Feb. 2019. 

\bibitem{emil_future2} L Sanguinetti, E. Bjornson, J Hoydis, ``Towards Massive MIMO 2.0: Understanding spatial correlation, interference suppression, and pilot contamination,'' arXiv:1904.03406 [eess.SP], 6 April 2019.


\bibitem{Weng} J. Weng, X. Tu, Z. Lai, S. Salous and J. Zhang, ``Indoor Massive MIMO Channel Modelling Using Ray-Launching Simulation,'' \textit{International Journal of Antennas and Propagation}, Art.-ID 279380, pp. 1-14, 2014. 


\bibitem{yevhen} Y. Yashchsyhyn, G. Bogdan, K. Godziszewski and P. Bajurko, ``Massive MIMO testing facility for future 5G systems,'' \textit{Proceedings of International Conference on Information and Telecommunication Technologies and Radio Electronics (UkrMiCo), Odessa}, pp. 1-5, 2017.


\bibitem{she} J. She, W. J. Lu, Y. Liu, P. F. Cui and H. B. Zhu, ``An Experimental Massive MIMO Channel Matrix Model for Hand-Held Scenarios,'' \textit{IEEE Access}, vol. 7, pp. 33881-33887, Mar 2019.


\bibitem{Mikki_book} S. Mikki and Y.  Antar, \textit{New Foundations for Applied Electromagnetics: The Spatial Structure of Fields}, Artech House, London, 2016.

\bibitem{mikki_cgf_2015} S. M. Mikki and Y. M. M. Antar, ``On Cross Correlation in Antenna Arrays With Applications to Spatial Diversity and MIMO Systems,'' \textit{IEEE Transactions on Antennas and Propagation}, vol. 63, no. 4, pp. 1798-1810, 2015.

\bibitem{clauzier3} S. Clauzier, S. M. Mikki, and Y. M. M. Antar, ``A Generalized Methodology for Obtaining Antenna Array Surface Current Distributions With Optimum Cross-Correlation Performance for MIMO and Spatial Diversity Applications,'' \textit{IEEE Antennas and Wireless Propagation Letters}, vol. 14, pp. 1451-1454, 2015.

\bibitem{mikki_access} S. M. Mikki, S. Clauzier and Y. M. M. Antar, ``Empirical Geometrical Bounds on MIMO Antenna Arrays for Optimum Diversity Gain Performance: An Electromagnetic Design Approach,'' \textit{IEEE Access}, vol. 6, pp. 39876-39894, 2018.

\bibitem{idm_cgf_mmimo_2019} D. Sarkar, S. M. Mikki and Y. M. M. Antar, ``Eigenspace Structure Estimation for Dual-Polarized Massive MIMO Systems Using an IDM-CGF Technique,'' \textit{IEEE Antennas and Wireless Propagation Letters}, vol. 18, no. 4, pp. 781-785, 2019.

\bibitem{yee} K. S. Yee, ``Numerical solution of initial boundary value problems involving Maxwell’s equations in isotropic media,'' \textit{IEEE Transactions on Antennas and Propagation}, vol. AP-14, no. 3, pp. 302-307, 1966.


\bibitem{fdtd_jensen_wallace1} J. W. Wallace and M. A. Jensen, ``Validation of parameteric directional MIMO channel models from wideband FDTD simulations of a simple indoor environment,'' \textit{Proceedings of  IEEE Antennas and Propagation Society International Symposium,} pp. 535-538, 2003. 

\bibitem{massive_MIMO_FDTD_RT} S. Shikhantsov, A. Thielens, G. Vermereen, E. Tanghe, P. Demeester, L. Martens, G. Torfs and W. Joseph, ``Hybrid Ray-Tracing/FDTD Method for Human Exposure Evaluation of a Massive MIMO Technology in an Industrial Indoor Environment,'' \textit{IEEE Access}, vol. 7, pp. 21020-21031, 2019.

\bibitem{ds_tap_1} D. Sarkar and K. V. Srivastava, ``Application of Cross-correlation Green’s Function along with FDTD for Fast Computation of Envelope Correlation Coefficient over Wideband for MIMO Antennas,'' \textit{IEEE Trans. on Antennas and Propagation}, vol. 65, no. 2, pp. 730-740, 2017.

\bibitem{tap_idm} D. Sarkar, S. M. Mikki, K. V. Srivastava and Y. M. M. Antar, ``Dynamics of Antenna Reactive Energy Using Time Domain IDM Method,'' \textit{IEEE Transactions on Antennas and Propagation}, vol. 67, no. 2, pp. 1084-1093, 2019.

\bibitem{nf_corr_awpl} S. M. Mikki, D. Sarkar and Y. M. M. Antar, ``Near-Field Cross-Correlation Analysis for MIMO Wireless Communications,'' \textit{IEEE Antennas and Wireless Propagation Letters}, vol. 18, no. 7, pp. 1357-1361, May 2019. 

\bibitem{Hu_2} S. Hu, F. Rusek, and O. Edfors, ``Beyond massive MIMO: The potential of data transmission with large intelligent surfaces,'' \textit{IEEE Transactions on Signal Processing}, vol. 66, no. 10, pp. 2746-2758, 2018.


\bibitem{Jung1} M. Jung, W. Saad, Y. Jang, G. Kong, S. Choi, ``Performance analysis of large intelligent surfaces (LISs): Asymptotic data rate and channel hardening effects,'' arXiv:1810.05667, Feb. 2019.

\bibitem{Nadeem} Q. U. A. Nadeem, A. Kammoun, A. Chaaban, M. Debbah and M. S. Alouini, ``Large Intelligent Surface Assisted MIMO Communications,'' arXiv:1903.08127v1, Mar. 2019.





\bibitem{Basar1} E. Basar \textit{et al}, ``Wireless Communications Through Reconfigurable Intelligent Surfaces,'' \textit{IEEE Access}, vol. 7, pp. 116753-116773, Aug 2019.

\bibitem{emil_irs} E. Bjornson, O. Ozdogan, E. G. Larsson, ``Intelligent Reflecting Surface vs. Decode-and-Forward: How Large Surfaces Are Needed to Beat Relaying?,'' arXiv:1906.03949v1, Jun 2019. 

\bibitem{ntontin} K. Ntontin \textit{et al}, ``Reconfigurable Intelligent Surfaces vs. Relaying: Differences, Similarities, and Performance Comparison,'' arXiv:1908.08747v1, Aug 2019. 

\bibitem{Hum} S. V. Hum and J. P. Carrier, ``Reconfigurable reflectarrays and array lenses for dynamic antenna beam control: A review,'' \textit{IEEE Transactions on Antennas and Propagation}, vol. 62, no. 1, pp. 183-198, 2014.


\bibitem{akbari} M. Akbari, F. Samadi, A. R. Sebak and T. A. Denidni, ``Superbroadband Diffuse Wave Scattering Based on Coding Metasurfaces,'' \textit{IEEE Antennas and Propagation Magazine}, vol. 61, no. 2, pp. 40-52, 2019. 

\bibitem{mmimo_FDTD_FSS_awpl} D. Sarkar, S. M. Mikki and Y. M. M. Antar, ``Engineering the Eigenspace Structure of Massive MIMO Links Through Frequency-Selective Surfaces,'' \textit{IEEE Antennas and Wireless Propagation Letters}, 2019.


\bibitem{bayatpur}  F. Bayatpur and K. Sarabandi, ``A tunable metamaterial frequency selective surface with variable modes of operation,'' \textit{IEEE Transactions on Microwave Theory and Techniques}, vol. 57, no. 6, pp. 1433–1438, Jun. 2009.


\bibitem{sghosh} S. Ghosh and S. Lim, ``Fluidically-Reconfigurable Multifunctional Frequency Selective Surface with Miniaturization Characteristic,'' \textit{IEEE Transactions on Microwave Theory and Techniques}, vol. 66, no. 8, pp. 3857-3865, 2018.




  






\end{thebibliography}
\end{document}